\documentclass{PoS}

\usepackage{graphicx}
\usepackage{hepunits}
\usepackage{slashed}

\newcommand{\msbar}{{\overline{\rm MS}}}

\newcommand{\bare}{{\rm bare}}

\newcommand{\Nf}{N_{\mathrm f}}
\newcommand{\pcac}{{\rm PCAC}}
\newcommand{\LambdaQCD}{\Lambda_{\rm QCD}}

\title{$N_f=1+2$ mass dependence of the topological susceptibility}

\ShortTitle{$N_f=1+2$ mass dependence of the topological susceptibility}

\author{\speaker{Julien Frison}\\
        KEK Theory Center, 1-1 Oho, 305-0801 Tsukuba, Japan\\
        E-mail: \email{jfrison@post.kek.jp}}

\author{Ryuichiro Kitano\\
        KEK Theory Center, 1-1 Oho, 305-0801 Tsukuba, Japan\\
        E-mail: \email{ryuichiro.kitano@kek.jp}}
\author{Norikazu Yamada\\
        KEK Theory Center, 1-1 Oho, 305-0801 Tsukuba, Japan\\
        E-mail: \email{norikazu.yamada@kek.jp}}

\abstract{A massless up quark has long been proposed as a solution to the strong CP problem. While this solution is sometimes thought to have been excluded, it
is actually still ill-defined. In this work, we study the mass dependence of the physical observable $\chi_t$, the topological susceptibility.
Assigning an unphysically large value to the down mass allows to be more sensitive to the non-perturbative effects behind the $m_u=0$ ambiguity. Preliminary results are presented for four masses of clover fermions. }

\FullConference{34th annual International Symposium on Lattice Field Theory\\
		24-30 July 2016\\
		University of Southampton, UK}

\begin{document}

\section{Introduction}

If one naively selects all the lagrangian terms allowed by symmetries, the QCD lagrangian should contain a CP-violating term $\theta F\tilde F$. The absence of experimental evidence for a non-zero $\theta$ has not been theoretically explained yet, and it is called strong CP problem. Several candidates exist to propose a solution to this problem, from the Peccei-Quinn mechanism to the ``trivial'' solution $m_u=0$. In this work we focus on the latter, arguing that it is not as trivial as it has been thought. 

Indeed, it has been previously\cite{Creutz:2007yr,Creutz:2010ts,Dine:2014dga} pointed out that $m_u=0$ is not renormalisation invariant, because of non-perturbative contributions. Even if $m_u=0$ for one particular scheme at some scale, it doesn't mean it will be zero for every scheme, and except for some particular schemes it doesn't mean that it will be zero at every scale. Only $m_u=m_d=0$ can be called physical, because it corresponds to a massless pion. 

In the past years, several lattice groups have estimated the quark masses at high precision in massless renormalisation schemes at perturbative scales, strongly excluding $m_u(\msbar)=0$. There the mass running is multiplicative and asymptotically universal, so that if $m_u$ were zero at some high energy scale it would be zero at other high energy scales as well. On the other hand, in instanton computations the quark masses typically appear renormalised at low energy, so that the $\msbar$ results are of limited use.

\section{Zero up mass or zero topological susceptibility?}

The idea of the $m_u=0$ solution of the strong $CP$ problem is that a $\theta F\tilde F$ term can be absorbed by a chiral rotation and is equivalent to putting a 
complex phase to $m_u$. Then, if $m_u=0$, $\theta$ becomes irrelevant since $m_u e^{i\theta}=0$ for any $\theta$.

If we start from $\chi_t=0$, where $\chi_t = \sum\langle (F\tilde F)(x)(F\tilde F)(0) \rangle/V$ is the topological susceptibility, we arrive to the same conclusion. Indeed it would impose that for every configuration we have the topological charge $Q=0$, and then the reweighting factor $e^{i\theta Q}$ from $\theta=0$ vacuum to arbitrary $\theta$ would be $1$ for any $\theta$. 

Those two things are equivalent at leading order in ChPT since $\chi_t \sim \Sigma m_u$. However, it has been known that the definition of $m_u$ is ambiguous at next to leading order\cite{Kaplan:1986ru}. The fact that $\chi_t(m_u=0,m_d=0)=0$ has also been checked empirically in full QCD on the lattice\cite{Hart:2001pj}, and in that case it makes sense that the massless pion allows to reproduce the ChPT result. But it is not clear whether the lattice results of full QCD respect $\chi_t(m_u=0)=0$ when up and down quark masses are not degenerate. It certainly depends on how we regularise the theory non-perturbatively and how we define the quark masses.

A crucial difference between those two approaches is that $\chi_t$ is a physical quantity, while $m_u$ is a non-observable parameter which depends on the renormalisation scheme. It then appears that the ``$m_u=0$ solution to the strong $CP$ problem'' would probably have been better expressed as a ``$\chi_t=0$ solution to the strong $CP$ problem''\footnote{The axion solution also provides $\chi_t=0$, but here we are interested in whether $\chi_t=0$ can be realised within QCD alone, under the constraint of reproducing the hadron spectrum}, and that lattice studies could focus on whether $\chi_t=0$ instead of focusing on a scheme-dependent quantity which is actually more difficult to obtain.  

\section{Renormalisation mixing and Ward identities}

The fact that the mass operator and the $F\tilde F$ operator mix is obvious in the singlet Axial Ward identity. This also puts a constraint relating the mass operator, the additive mass renormalisation and the $F\tilde F$ operator. Then taking an arbitrary definition for each term is likely to break the Ward identity, unless the observables are explicitely defined from it\cite{Bochicchio:1984uv}. 

Since $F\tilde F$ is related to instantons, it is natural to expect that the 't Hooft vertex could play a role in this mixing. In $\Nf=1$ the t' Hooft vertex would be a two leg vertex giving a $O(\LambdaQCD)$ contribution. For larger $\Nf$ the t'Hooft vertex has $2\Nf$ legs, of which the $2\Nf-2$ legs associated to ``heavy'' quarks can be contracted with their mass terms so that in $\Nf=3$ the up quark appears to receive a contribution $\Delta m_u = O(\frac{m_dm_s}{\LambdaQCD})$. 

It has been estimated\cite{Dine:2014dga} that those corrections could be large at an $O(\LambdaQCD)$ renormalisation scale. A nonsignificant effect on the lattice results for $m_u(\msbar,3\ \GeV)$ could become a large (scheme-dependent) effect at low-energy scales. 

Therefore, the question of whether $\chi_t(m_u=0)$ cancels could be dramatically affected by the choice of definition of the quark masses. In particular we are interested in whether $\chi_t=0$ at $m_u^\pcac=0$, since the ``PCAC mass'' from non-singlet Ward identities is commonly used to extract the $\msbar$ masses from lattice calculations. 

\section{Comparison of topological charge definitions}

As a first step, we focused on understanding the difference between different topological charge observable definitions. Indeed, just like $\chi_t(m_u)$ could depend of the definition of $m_u$ it could depend on the definition of $\chi_t$. One could in particular expect differences between the fermionic and bosonic definitions of $\chi_t$, in particular with lattice actions which don't satisfy the index theorem. It was also argued that some definitions were more compatible with the Axial Ward identities than others, which could be important since we're interested in $m_u^\pcac$.

\begin{figure}\begin{center}
  \includegraphics[width=0.3\linewidth]{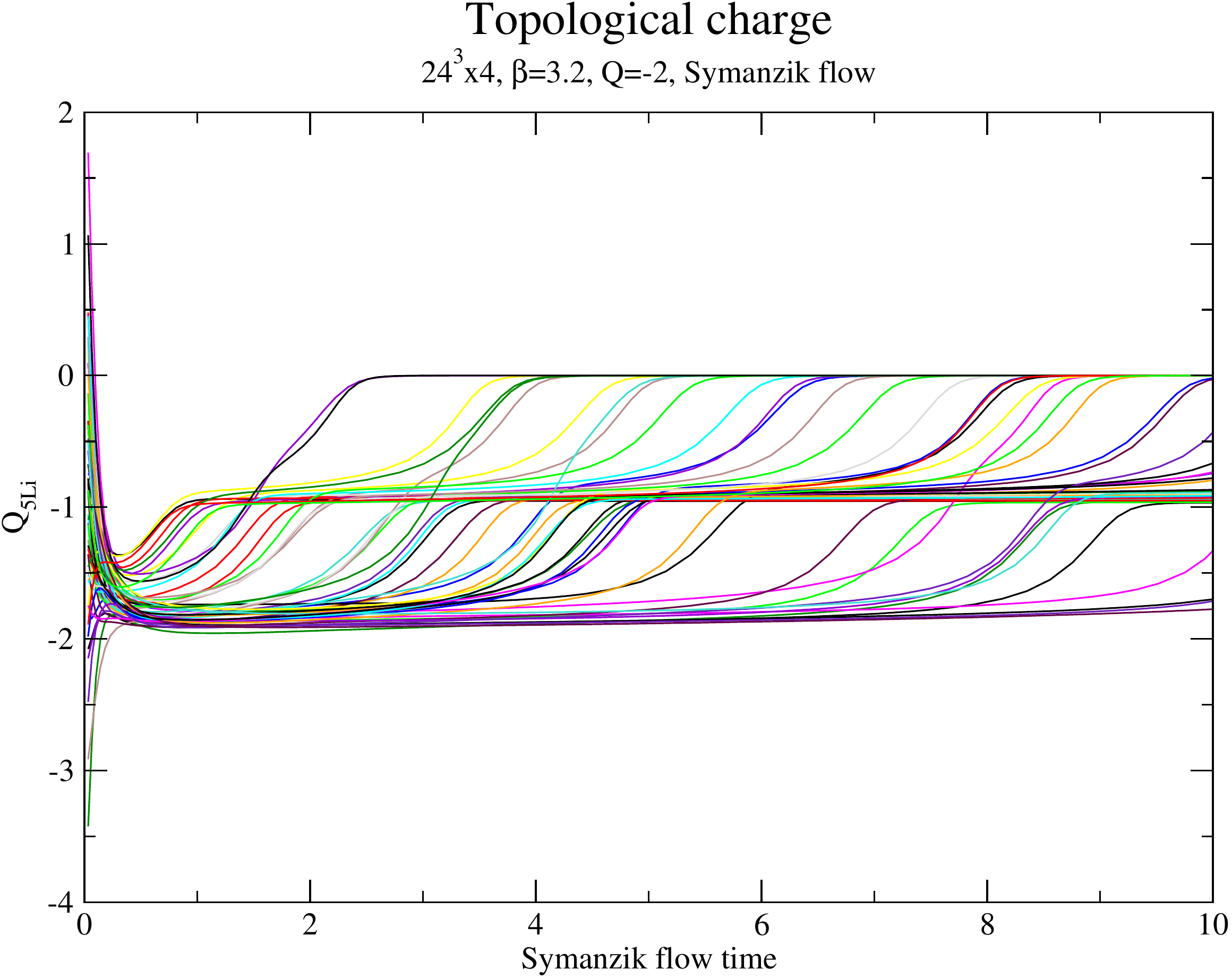}
  \hfill
  \includegraphics[width=0.3\linewidth]{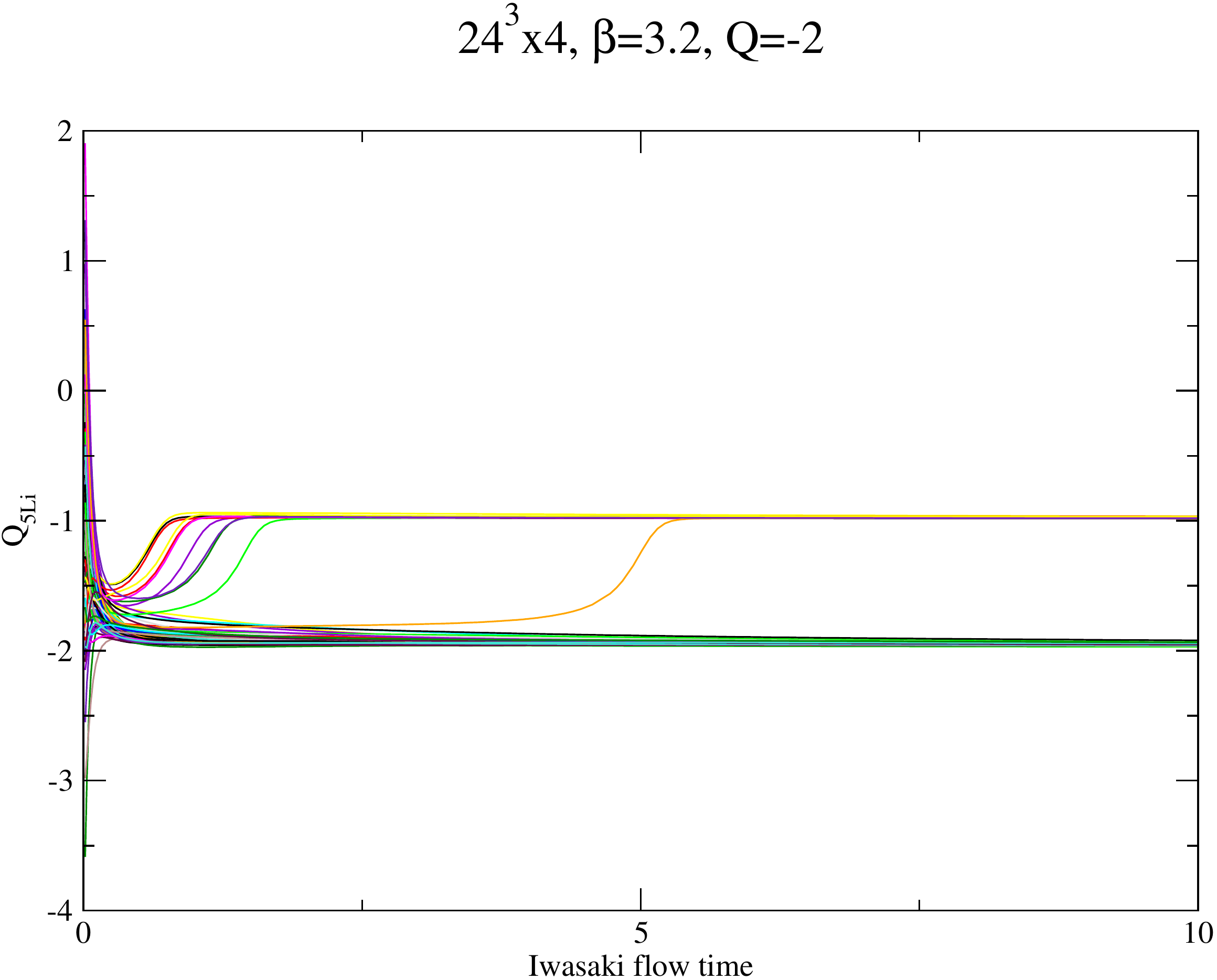}
  \hfill
  \includegraphics[width=0.3\linewidth]{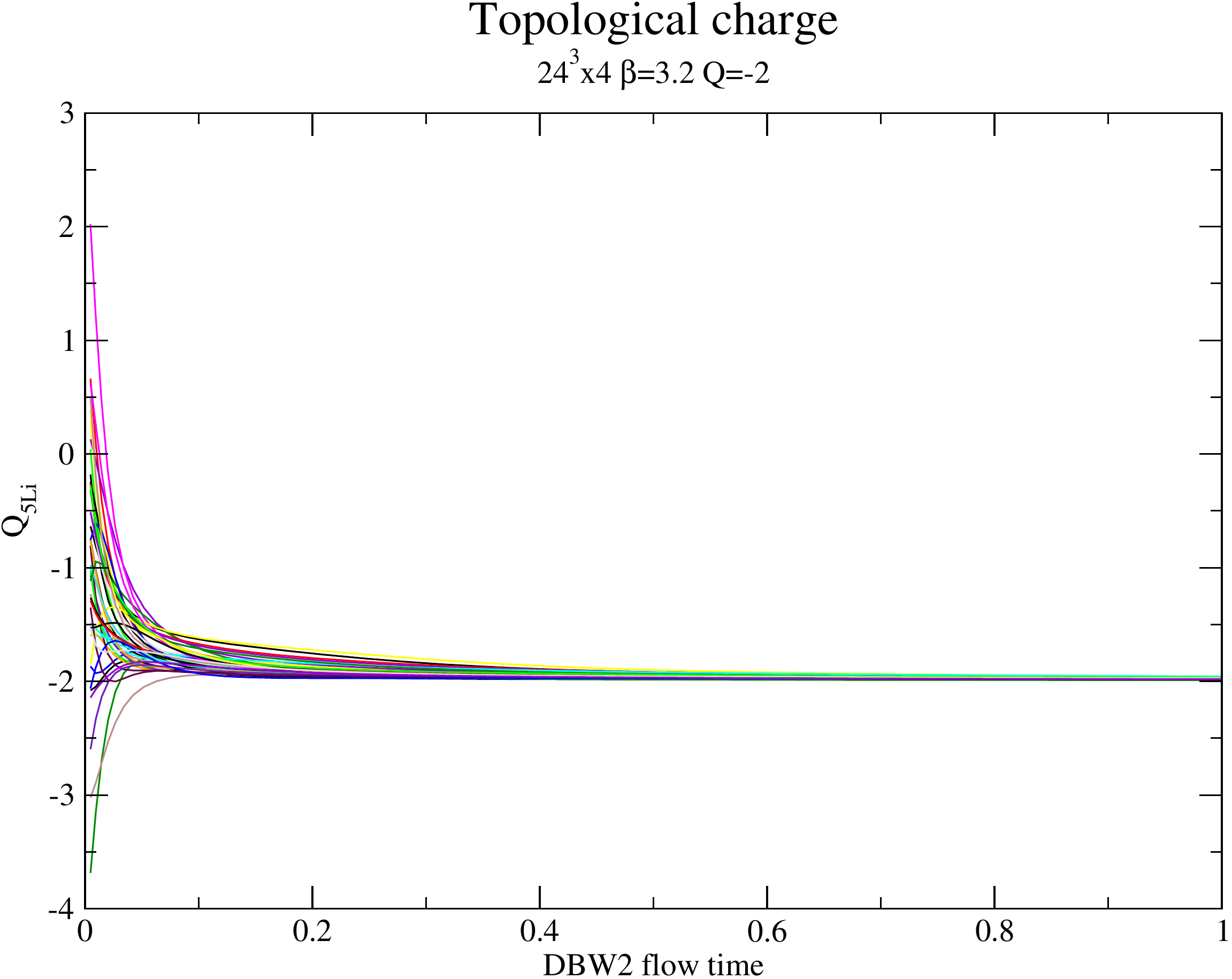}
  \caption{Gradient flow evolution with different choices of action, for a small Iwasaki quenched ensemble with a topology-fixing term in the action. Every line represents a different lattice configuration.}
  \label{fig:tcharge_24x4_Q-2}
\end{center}\end{figure}

In Fig.~\ref{fig:tcharge_24x4_Q-2} we show how gradient-flow definitions of $Q$ with different actions compare, on a ensemble which is generated in a fixed topological sector of $Q=-2$. This topology fixing uses a fermionic definition, suppressing zero modes, while the measurement after gradient flow uses a $5Li$ gluonic definition. Although the stability of the plateaus depends very much on the choice of flow action, we see that the DBW2-flow perfectly agree with the fermionic definition giving $Q=-2$. As a general remark, increasing $c_1$ in the action seems to increase both the stability and the convergence speed (so that the equivalent number of smearing steps is roughly constant, if we use $n_c=(3-15c_1)\tau$ from \cite{Alexandrou:2015yba}).

\begin{figure}\begin{center}
  \includegraphics[width=0.45\linewidth]{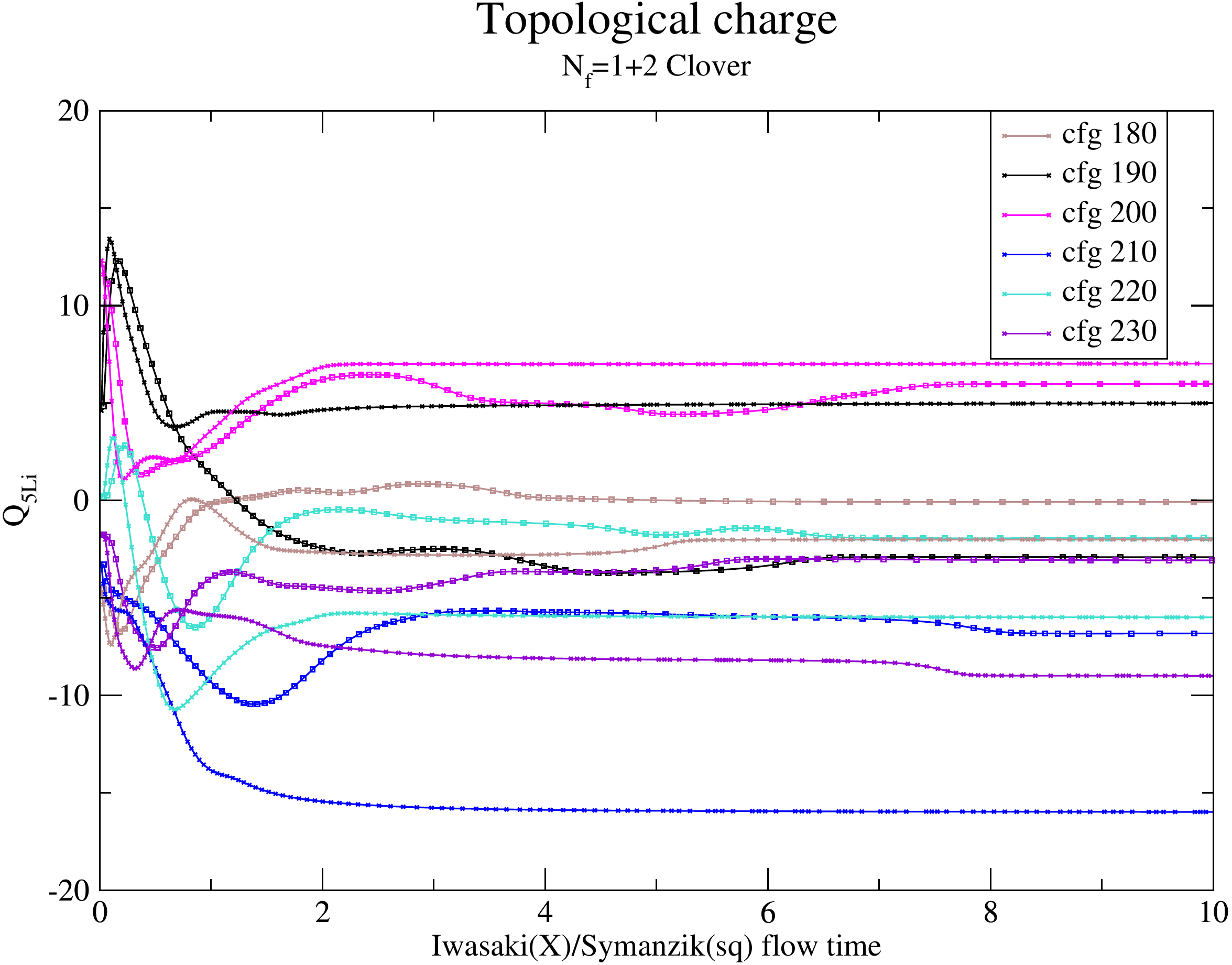}
  \hfill
  \includegraphics[width=0.45\linewidth]{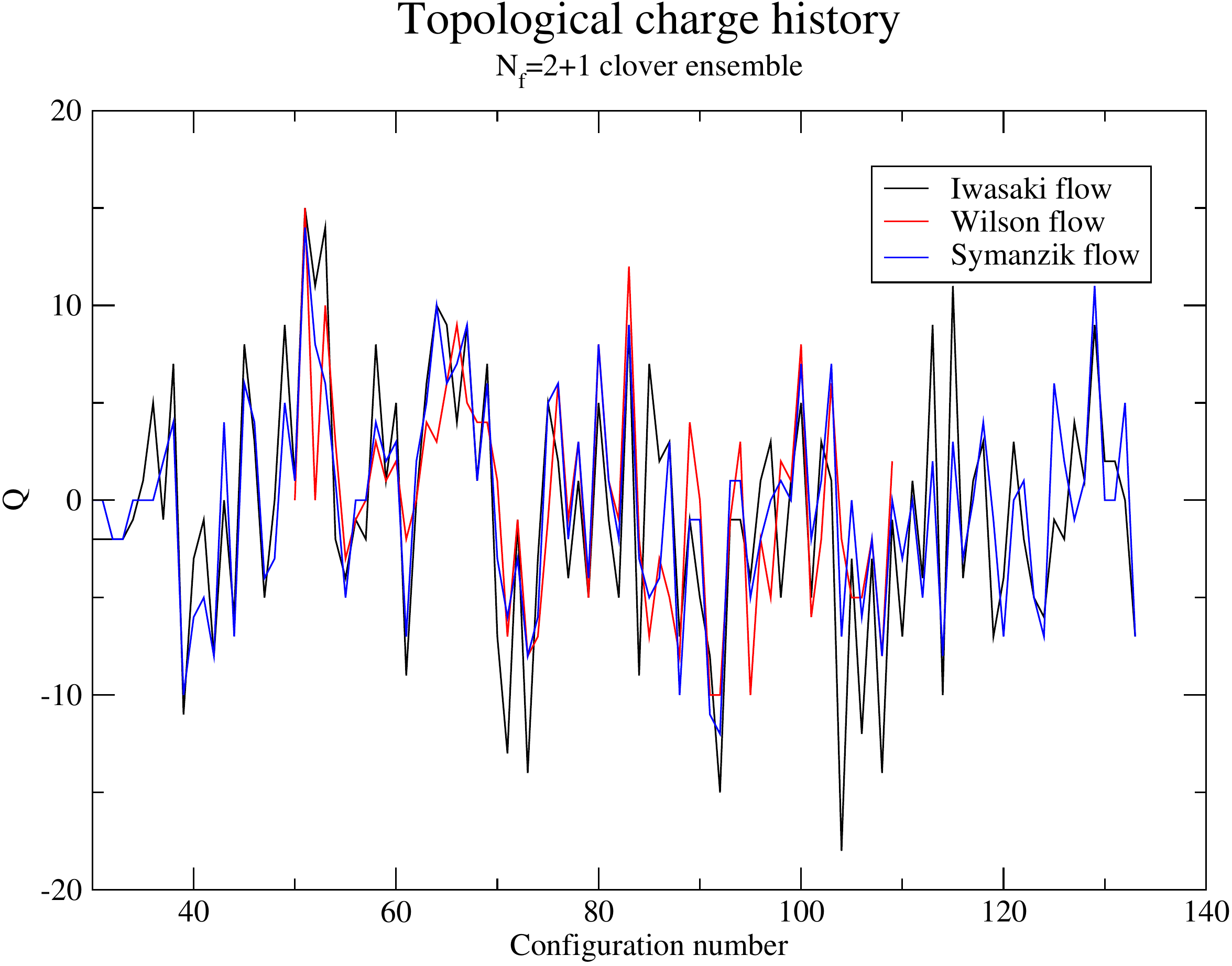}
  \caption{In the left panel we show that topological charge definitions coming from two different actions seem to completely disagree. For one configuration taken independently there is no asymptotic agreement between Symanzik flow (square symbol) and Iwasaki flow (cross). However, the right panel, presenting the same information at fixed flow time in a different way, exhibits a strong correlation between the different definitions. }
  \label{fig:tcharge_Nf1p2}
\end{center}\end{figure}

\begin{figure}\begin{center}
  \includegraphics[width=0.6\linewidth]{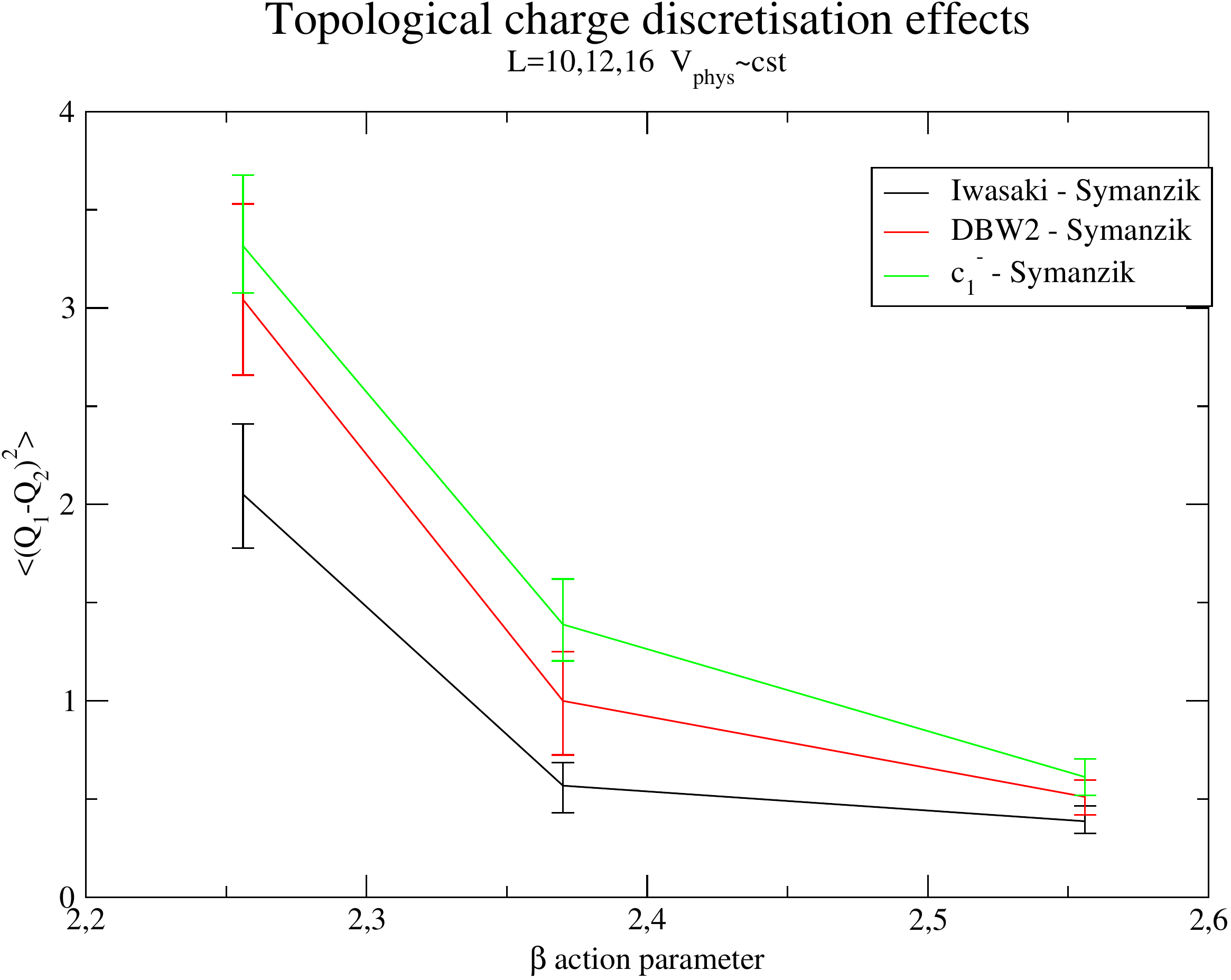}
  \caption{For Iwasaki gauge action ensembles where the physical volume and $\chi_tV\sim 13$ are kept approximatively constant, we see that the discrepancy between different definitions of the topological charge based on different flows disappears in the continuum (high $\beta$). $c_1^-$ stands for an action with the rectangle coefficient $c_1=-10$. }
  \label{fig:Q2_cont}
\end{center}\end{figure}

In Fig.~\ref{fig:tcharge_Nf1p2}-\ref{fig:Q2_cont} we show other tests to check that all definitions are reasonably consistent within our range of parameters, at least in the case of quenched approximation or unphysically large quark masses. We hope to eventually perform those comparisons on dynamical configurations with near-zero up quark mass. 

\section{$\Nf=1+2$ simulation}

As we want to highlight the additive mass contribution to $m_u$, we make the choice to generate $\Nf=1+2$ ensembles, where the up quark is light but the the down quark is taken degenerate with the strange and close to the physical strange mass. This will put us close to the $\Nf=1$ case where non-perturbative effects are supposed to add an $O(\LambdaQCD)$ additive contribution to renormalisation, and this will enhance any term in $O(\frac{m_dm_s}{\LambdaQCD})$.

We use a L\"uscher-Weisz gauge action with HEX2-smeared clover fermions. We hereby present preliminary results for one single coupling $\beta=3.31$ (corresponding to $a\sim 0.116\ {\rm fm}$ in the $\Nf=2+1$ scale-setting) and lattice size $16^3\times 32$. The down/strange quark mass is kept fixed at $m_s^\bare=-0.04$ while $m_u^\bare$ varies from $-0.07$ to $-0.1$. $O(100)$ configurations have been generated for each mass. 

\begin{figure}\begin{center}
  \includegraphics[width=0.8\linewidth]{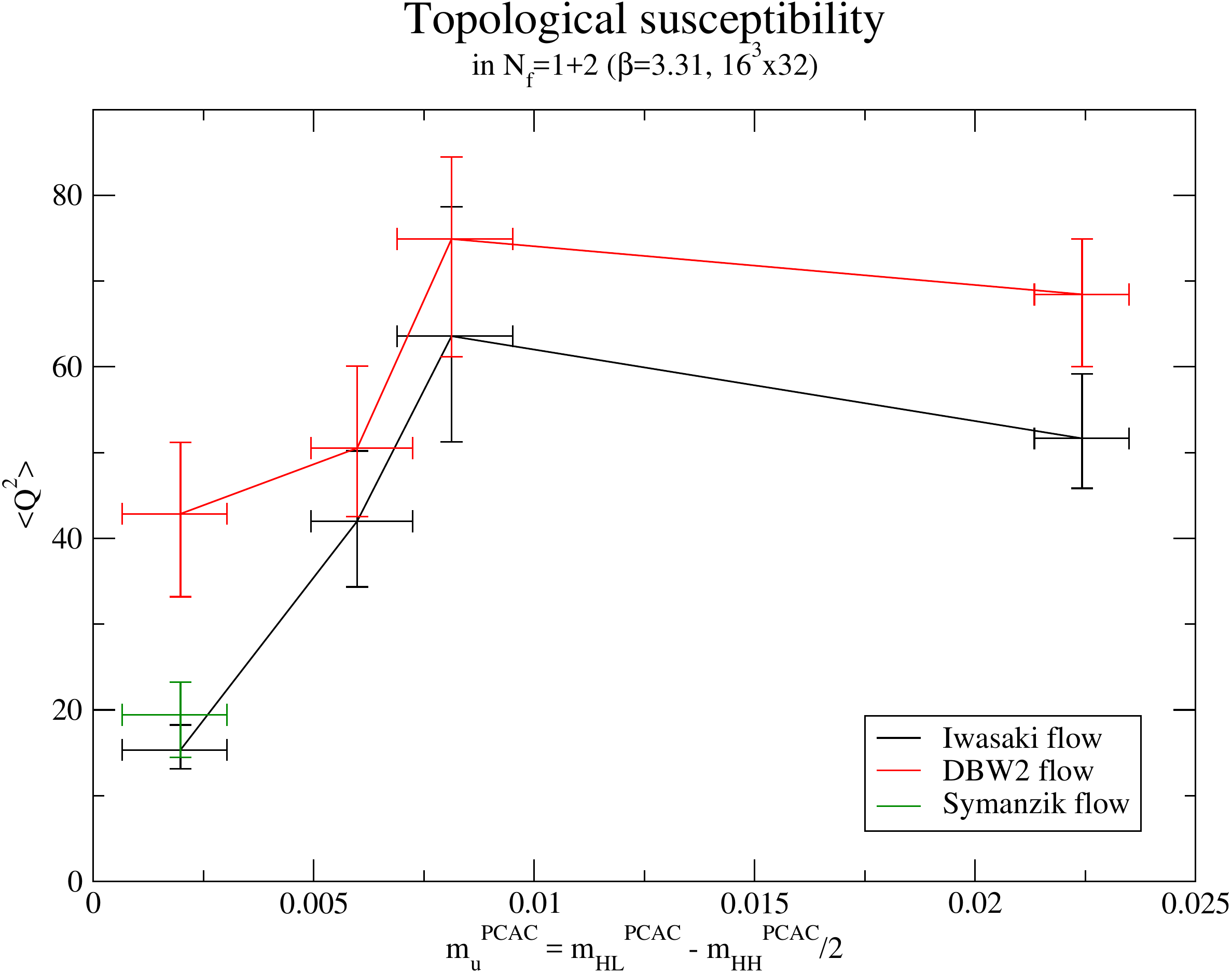}
  \caption{Topological charge as a function of the up quark PCAC mass, at a fixed volume and lattice spacing. The ``down'' and strange mass are kept constant and close to the physical strange mass. }
  \label{fig:Q2_vs_PCAC}
\end{center}\end{figure}

The PCAC masses are extracted from the heavy-light and heavy-heavy non-singlet Axial Ward identity and then combined into
\begin{equation}
  m_u^\pcac = m_{HL}^\pcac - m_{HH}^\pcac. 
\end{equation}
This is presented in Fig.~\ref{fig:Q2_vs_PCAC} together with gradient flow definitions of the topological susceptibility. 

The topological susceptibility at large quark mass looks more or less flat and is comparable with the quenched value. For lighter quark masses we see a decrease of the topological susceptibility, as expected, but those preliminary results do not yet allow to tell whether $\chi_t=0$ at $m_u^\pcac=0$ or not. 

\section{Conclusion}

We have argued that the possibility that existing determinations of $m_u$ in perturbative or massless schemes do not allow to properly define and exclude the $m_u=0$ scenario of the strong CP problem. To our knowledge, there is no theoretical proof that $\chi_t$ cancels when one of the PCAC masses does. We have therefore presented preliminary results for an empirical test of this property on $\Nf=1+2$ ensembles when the up and down quarks are non-degenerate. 

The results still contain several sources of systematics which are not yet totally under control, and more ensembles are currently being generated. The continuum limit in particular is going to be an important ingredient. The statistics will also have to be improved by a large factor, while some standard analysis methods such as stochastic propagators can be used to improve our determination of the PCAC mass. 

Unfortunately, we have not been able to perform many analysis with a fermionic definition of the topological charge as a comparison, due to the heavy cost of overlap fermions. 

However, it appears that we can explore the up quark mass regime lighter than expected, probably due to the dynamics of $\Nf=1+2$ flavour QCD, {\it i.e. absence of light pions}. Thus, it is quite encouraging that we need not rely on a long extrapolation. We might even end up {\it interpolating} to $m_u^\pcac=0$.

\section*{Acknowledgements}
This work is supported by JSPS KAKENHI Grant-in-Aid for Scientific 
Research (B) (No. 15H03669 [RK]), MEXT Grant-in- Aid for Scientific 
Research on Innovative Areas (No. 25105011 [RK]) and the Large Scale 
Simulation Program No. 15/16-21 of High Energy Accelerator Research 
Organization (KEK).


\begin{thebibliography}{99}
\bibitem{Creutz:2007yr}
    M.~Creutz,
        Annals Phys.\  {\bf 323} (2008) 2349
	    [arXiv:0711.2640 [hep-ph]].
	  \bibitem{Creutz:2010ts}
	      M.~Creutz,
		  Phys.\ Rev.\ D {\bf 83} (2011) 016005
		      [arXiv:1010.4467 [hep-ph]].
		    \bibitem{Dine:2014dga}
		        M.~Dine, P.~Draper and G.~Festuccia,
			    Phys.\ Rev.\ D {\bf 92} (2015) no.5,  054004
			        [arXiv:1410.8505 [hep-ph]].

			      \bibitem{Hart:2001pj}
				  A.~Hart {\it et al.} [UKQCD Collaboration],
				      Phys.\ Lett.\ B {\bf 523} (2001) 280
					  [hep-lat/0108006].

					\bibitem{Bochicchio:1984uv}
					    M.~Bochicchio,
					        Nucl.\ Phys.\ B {\bf 271} (1986) 698.

					      \bibitem{Alexandrou:2015yba}
						  C.~Alexandrou, A.~Athenodorou and K.~Jansen,
						      Phys.\ Rev.\ D {\bf 92} (2015) no.12,  125014
							  [arXiv:1509.04259 [hep-lat]].

							\bibitem{Kaplan:1986ru}
							    D.~B.~Kaplan and A.~V.~Manohar,
							        Phys.\ Rev.\ Lett.\  {\bf 56} (1986) 2004.
								  doi:10.1103/PhysRevLett.56.2004

\end{thebibliography}
\end{document}